\documentclass[aps,prl,twocolumn,a4paper]{revtex4}

\usepackage{graphicx}
\usepackage{ulem}
\usepackage{amssymb}

\begin{document}


\title{Comment on ``New Scaling of Child-Langmuir Law in the Quantum Regime''}
 


\author{Debabrata Biswas} 

\affiliation{Theoretical Physics Division,
Bhabha Atomic Research Centre,
Mumbai 400 085, INDIA}






\maketitle

\newcommand{\be}{\begin{equation}}
\newcommand{\ee}{\end{equation}}
\newcommand{\bea}{\begin{eqnarray}}
\newcommand{\eea}{\end{eqnarray}}
\newcommand{\Tbar}{{\bar{T}}}
\newcommand{\En}{{\cal E}}
\newcommand{\Lop}{{\cal L}}
\newcommand{\DB}[1]{\marginpar{\footnotesize DB: #1}}
\newcommand{\q}{\vec{q}}
\newcommand{\kt}{\tilde{k}}
\newcommand{\Lopn}{\tilde{\Lop}}
\newcommand{\noi}{\noindent}
\newcommand{\ovn}{\bar{n}}
\newcommand{\ovx}{\bar{x}}
\newcommand{\ovE}{\bar{E}}
\newcommand{\ovV}{\bar{V}}
\newcommand{\ovU}{\bar{U}}
\newcommand{\ovJ}{\bar{J}}
\newcommand{\calE}{{\cal E}}



In their letter on the quantum Child-Langmuir law, Ang et al \cite{ang2003} include 
exchange correlation effects within the 
Kohn-Sham density functional theory (DFT) \cite{KS}
and explore numerically the maximum transmitted current ($J_{max}$) in nanogaps.
We show here that the calculations are in error as the exchange-correlation component
of the chemical potential has been ignored while fixing the boundary conditions for
the Hartree potential.

The analysis in \cite{ang2003} is based on solving the time-independnent Schrodinger equation
$-d^2 \psi/dx^2 + V_{eff} \psi = E \psi$ with an effective potential energy, 
$V_{eff} = -eV + V_{xc}\times E_H$, where 
$E_H = e^2/(4\pi\epsilon_0 a_0)$ is the 
Hartree energy, $a_0$ the Bohr radius,
$V_{xc} = \epsilon_{xc} - (r_s/3) d\epsilon_{xc}/dr_s$, $r_s = [3/(4\pi n)]^{1/3}$ is 
the Wigner-Seitz radius, $n$ the electron 
number density and $e$ the magnitude of the electronic charge. In the above,
$\epsilon_{xc}$ is the exchange-correlation energy density within the local
density approximation, and 
$V$ is the Hartree potential satisfying the Poisson equation
$d^2 V/dx^2 = e n(x)/\epsilon_0 =  e |\psi(x)|^2/\epsilon_0$
with boundary conditions $V(0) = 0$ and $V(D) = V_g$ where $V_g$ is the gap
voltage difference and $D$ the gap size.
These equations are solved in a nanogap with the wavefunction $\psi$ and its
derivative matched at the collector boundary under certain assumptions
(Ref. \cite{epjb} addresses one of these).

We first note that the applied voltage difference, $V_g = -(\mu_C - \mu_E)/e$ 
where $\mu_C$ and
$\mu_E$ refer respectively to the chemical potential at the collector and
injection planes. For convenience, we consider the reference as
$\mu_E = -eV(0) + V_{xc}(0)\times E_H = 0$ so that $E = 0$ refers to injection from
the Fermi level. Thus
$V(0) = V_{xc}(0) \times E_H/e$. It follows that the chemical potential at the
collector is $-eV(D) + V_{xc}(D) \times E_H = -eV_g$. 
Thus $V(D) = V_g + V_{xc}(D) \times E_H/e$.
In writing the above, we have implicitly assumed continuity of
the chemical potential at the interfaces under steady-state conditions.  
Note that when exchange-correlation
is neglected altogether, the boundary conditions for $V$ are $V(0) = 0$ and $V(D) = V_g$
respectively as assumed in \cite{ang2003}.

\begin{figure}[t]
\begin{center}
\includegraphics[width=4cm,angle=270]{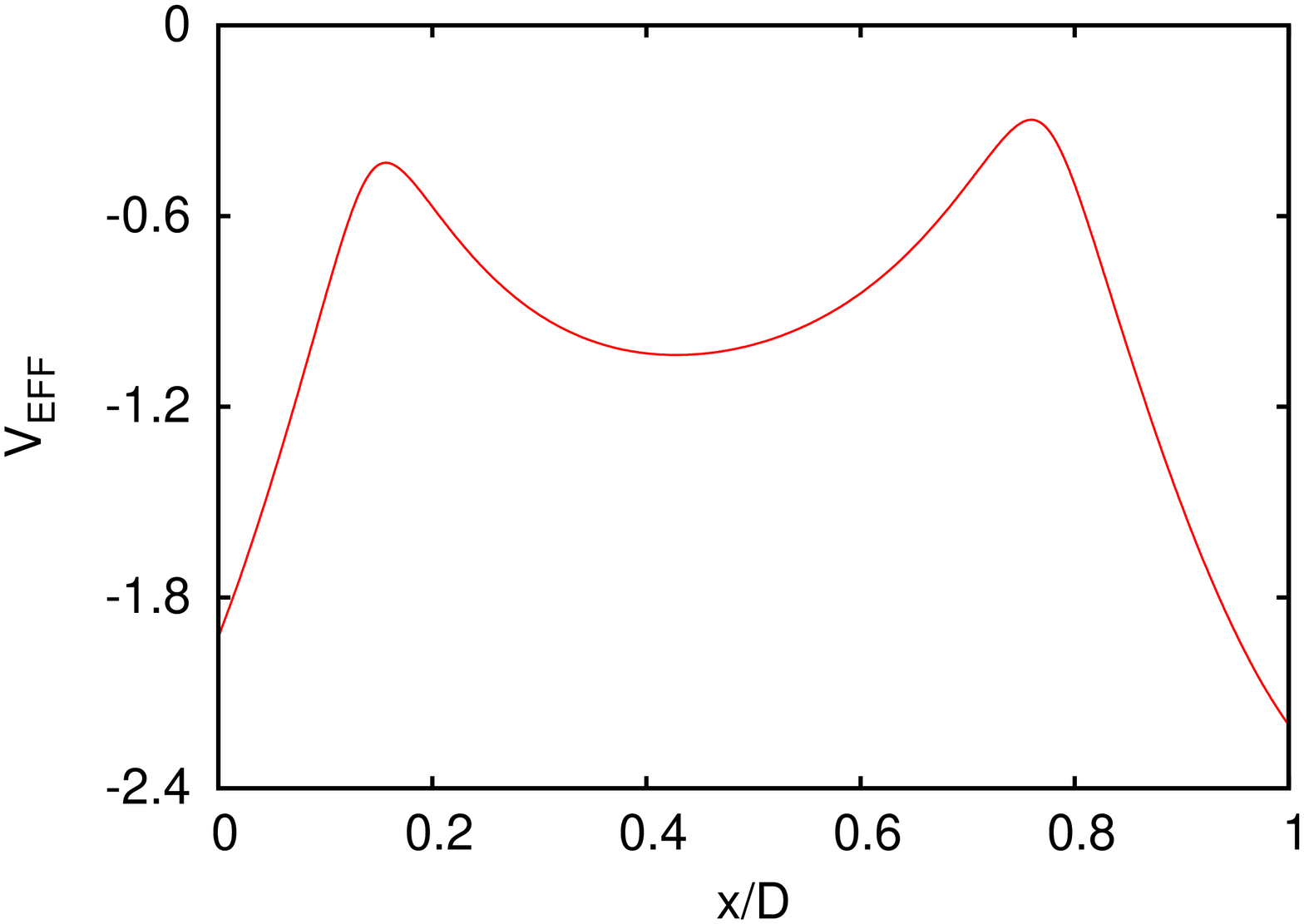}
\includegraphics[width=4cm,angle=270]{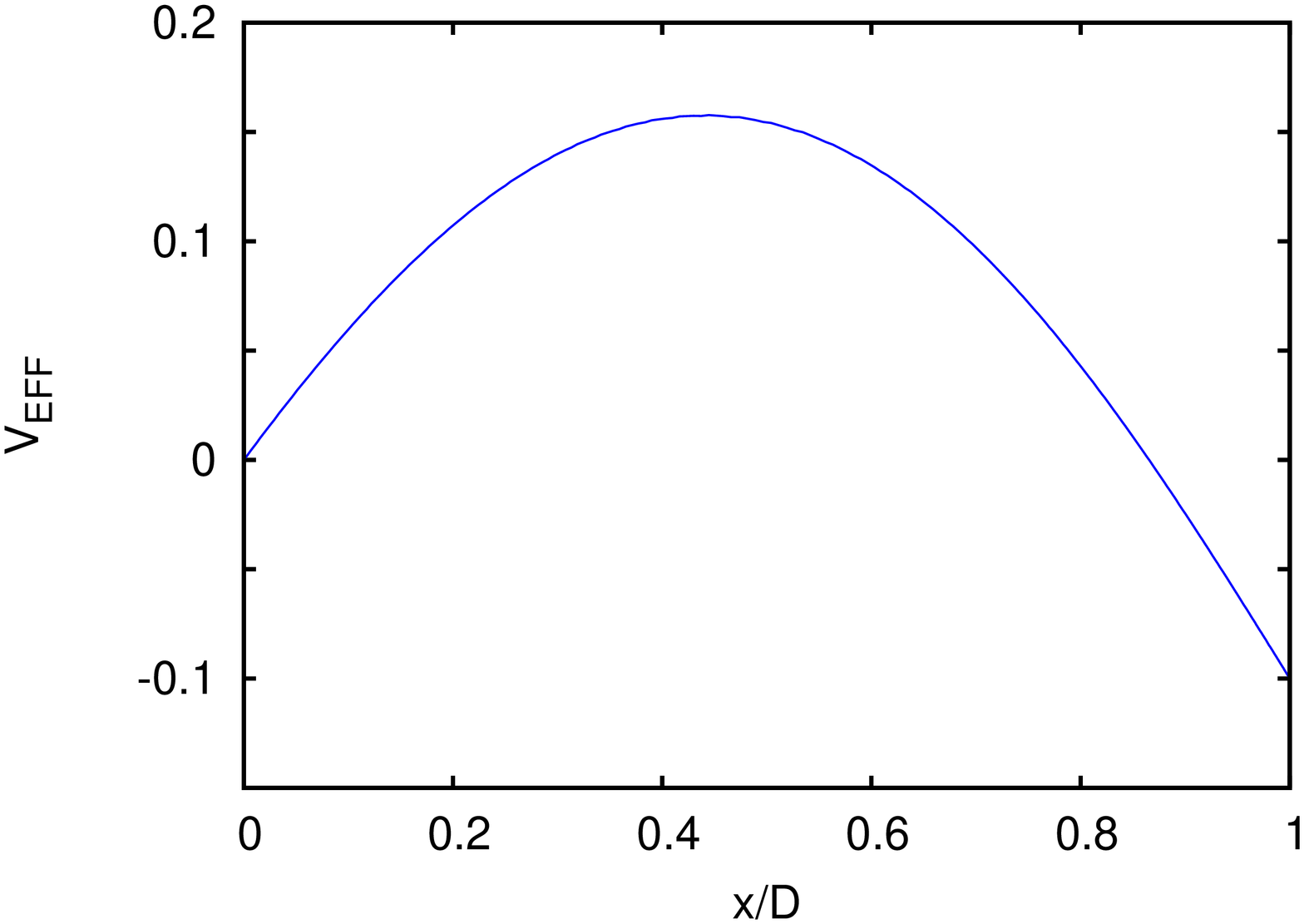}
\end{center}
\caption{The effective potential energy $V_{eff} = -eV + E_H\times V_{xc}$ for 
$D=1$nm, $V_g = 0.1$V ({\bf top}) $V(0) = 0$, $V(D) = V_g$ and $J_{max} \simeq 278 J_{CL}$
({\bf bottom}) $V(0) = V_{xc}(0) \times E_H/e$, $V(D) = V_g + V_{xc}(D) \times E_H/e$
and $J_{max} \simeq 3.4 J_{CL}$. The boundary conditions for $\psi$ are the
same as in Ref. \cite{ang2003} in both cases. }
\label{fig:Veff}
\end{figure}

When $V_{xc}$ is substantial, and the boundary
conditions chosen are $V(D) = V_g$ and $V(0) = 0$,
the results can be unphysical.
As an example,
consider the case $D = 1$nm, $V_g = 0.1$V and $E = 0$. Using the formalism 
of Ref. \cite{ang2003}, $J_{max}$ turns out
to be $J_{max} \simeq 278 J_{CL}$ where $J_{CL}$ is the classical Child-Langmuir
current density. The corresponding effective potential energy is shown in Fig.~\ref{fig:Veff}.
Note that $V_{eff}$ is negative everywhere. The injection energy $E$ has
clearly no relation to the effective potential at either end of the nanogap.

The corrected boundary conditions for $V$ lead to $J_{max} \simeq 3.4 J_{CL}$. The
corresponding effective potential is shown in Fig.~\ref{fig:Veff} (bottom).
The error in determining the quantum Child-Langmuir law is therefore substantial.
 
In regimes where $V_{xc}$ is negligible (large $V_g$ or $D$),  
$V(D) = V_g$ and $V(0) = 0$ are
approximately the correct boundary conditions within the formalism of Ref.~\cite{ang2003} 
as we have verified for $D = 50$nm, $V_g = 50$V and $E=0$.

Finally, we note that the error in boundary conditions for $V$ persists in subsequent 
publications by the authors of Ref. \cite{ang2003}. Results incorporating the rectified
boundary conditions for $V$ and improved boundary conditions for $\psi$ will
be published in a separate communication.

The author acknowledges fruitful discussions with Dr. Raghwendra Kumar and
Dr. Biplab Ghosh.


\end{document}